\documentclass[%
 reprint,
 amsmath,amssymb,
 aps,
]{revtex4-2}

\usepackage{graphicx}
\usepackage{dcolumn}
\usepackage{bm}
\usepackage{mathtools}
\usepackage{breqn}
\usepackage{physics}
\usepackage{caption}
\usepackage{subcaption}
\usepackage{subcaption}
\usepackage{listings}
\usepackage[section]{placeins}
\usepackage{mathrsfs}

\usepackage{xcolor}

\begin{document}

\preprint{APS/123-QED}

\title{Critical Three-Dimensional Ising Model on Spheriods from the Conformal Bootstrap\\}

\author{Daniel Berkowitz}
\affiliation{
 Yale University \\
 Department of Physics \\ email:daniel.berkowitz@yale.edu 
} 
\author{George T.\ Fleming}
\affiliation{
 Yale University \\
 Department of Physics \\ email:george.fleming@yale.edu 
}

\date{\today}

\begin{abstract}
We construct a conformal map from $\mathbb{R}^3$ to a three-dimensional spheriod, which
includes $\mathbb{S}^3$, a double-cover of the 3-ball, and $\mathbb{R} \times \mathbb{S}^2$
as limiting cases.  Using the data of the critical three-dimensional Ising model
on $\mathbb{R}^3$ that was computed using the conformal bootstrap method, we numerically
estimate the fourth-order Binder cumulant of the critical three-dimensional $\phi^4$ theory
on $\mathbb{S}^3$.  We expect this estimate will enable an interesting
comparison between the conformal bootstrap and future calculations of critical $\phi^4$ theory
on $\mathbb{S}^3$ using the Quantum Finite Element (QFE) method.

\end{abstract}

\maketitle


\section{\label{sec:level1}INTRODUCTION}
The last decade has seen major advances in the general study of conformal field
theories beyond the special cases of two-dimensional spacetimes or maximal
supersymmetry through the widespread development of the conformal boostrap
\cite{zamolodchikov1996conformal}.  Particularly notable is the success
of the conformal boostrap in constraining the data
(scaling dimensions and OPE coefficients) of the three-dimensional
critical Ising model CFT \cite{el2012solving, kos2014bootstrapping,
Kos:2016ysd, komargodski2017random},
surpassing
the previous best results using Markov chain Monte Carlo (MCMC)
on regular cubic discretizations of a three-dimensional torus
\cite{Hasenbusch:2010, Caselle:2015csa, Costagliola:2015ier}.
This has also driven the development
of improved methods for constructing three-dimensional conformal blocks,
the basis functions of the conformal group analogous to spherical harmonics
for the rotation group \cite{kos2014bootstrapping, penedones2016recursion,
hogervorst2016dimensional}.  These ingredients allow for an accurate
estimate of the four-point functions \cite{rychkov2017non}
of the critical Ising model.


In order to keep up with developments in the conformal bootstrap,
a new approach for MCMC-based calculations optimized for the study
of conformal fixed points in quantum field theories, called
Quantum Finite Elements (QFE), has recently been developed
\cite{Brower:2012vg, Brower:2012mn, Brower:2014daa, Brower:2015zea,
Brower:2016moq, Brower:2016vsl, Brower:2018szu, Brower:2020jqj}.
QFE is a general framework
for MCMC calculations on static curved manifolds.  Relevant to the study
of CFTs, there is no conformal map from flat Euclidean space $\mathbb{R}^3$,
where conformal bootstrap calculations are performed, to the torus
$\mathbb{T}^3$, where traditional MCMC calculations on cubic grids are performed,
making direct comparison in the infinite volume limit difficult for most
of the CFT data except a few leading scaling dimensions determined via
finite size scaling.  As we will show,
conformal maps from $\mathbb{R}^3$ to spheroids, including the special cases
of the sphere $\mathbb{S}^3$; the double cover of the 3-ball; and the
cylinder $\mathbb{R} \times \mathbb{S}^2$, do exist and MCMC calculations
can be performed on these manifolds using QFE.  In particular, observables
can be constructed in QFE to directly calculate all the CFT data that
appears in the conformal block expansion: scaling dimensions and OPE
coefficients.

As we shall see, one class of observables where the accuracy of QFE
calculations are expected to surpass those of the conformal bootstrap
is in the calculation of moments of the average magnetization
\begin{equation}
\begin{aligned}
&M = \int d^3x \sqrt{g(x)} \phi(x) , \quad \\&
m_n = \left\langle \int d^3x \sqrt{g(x)} \left( \phi(x) - M \right)^n
\right\rangle.
\end{aligned}
\end{equation}
In QFE and traditional calculations on cubic lattices, moments
of average magnetization can be computed very accurately due
to the increased statistics of averaging over the volume.

We will first define the manifolds
we are working with and obtain a general Weyl factor which will allow us to define
a conformal map between a 3D spheroid and $\mathbb{R}^{3}$ and discuss the limiting
cases of $\mathbb{S}^3$, $\mathbb{R} \cross \mathbb{S}^2$, and the double-cover
of 3-ball in detail.  We will show how to map the approximate four-point function
computed in the conformal bootstrap on $\mathbb{R}^3$ for the critical 3D Ising model
to the four-point function on a spheroid.  Then, we integrate the
two-point function and apply Monte-Carlo integration to the four-point function and obtain estimates for the fourth-order Binder
cumulant of the critical 3D Ising model on $\mathbb{S}^3$.  Finally, we will discuss
how this result can be compared with future lattice calculations, including
quantum finite elements (QFE).


\section{Conformal Invariance of a 3D Spheroid}
\subsection{Conformal Invariance of $\mathbb{S}^3$}

In this section we will furnish a conformal mapping between a general 3D spheroid and $\mathbb{R}^3$. We will later use the resultant Weyl factor to obtain an estimate of the fourth-order Binder cumulant for the critical 3D Ising model on $\mathbb{S}^3$. Our spheroid can be defined as a set of points embedded in $\mathbb{R}^4$ which satisfies the following relation in Cartesian coordinates 

\begin{equation}\label{1rrr}
\frac{x^{2}}{a^{2}}+\frac{y^{2}}{a^{2}}+\frac{z^{2}}{a^{2}}+\frac{w^{2}}{b^{2}}=1 \quad(a, b>0).
\end{equation}
When $b \rightarrow 0$ our spheroid approaches the 3D analogue of a disc or a 2-ball, which is a 3-ball. Technically speaking this 3-ball will be the superposition of two 3-balls superimposed onto each other at their boundaries. This will be discussed in more detail later. In the opposite limit when $b \rightarrow \infty$ our spheroid can be understood to approach $\mathbb{R} \cross \mathbb{S}^2$. In an earlier calculation $\phi^{4}$ theory at its Wilson-Fisher fixed point was analyzed \cite{Brower:2020jqj} on $\mathbb{R} \cross \mathbb{S}^2$ using the QFE. When $ b\rightarrow a$ the spheroid approaches $\mathbb{S}^3$, which is the next manifold we wish to study $\phi^{4}$ theory on using the QFE. 

A set of equations which satisfy (\ref{1rrr}) and encapsulates all of the cases we just outlined and everything else in-between are  

\begin{equation}\label{2rrr}
\begin{aligned}
&x=a \sin \psi \sin \theta \cos\phi, \\& y=a \sin \psi \sin \theta \sin \phi, \\& z=a \sin \psi \cos \theta,  \\& w= b \cos \psi, 
\end{aligned}
\end{equation}
where $\psi$ and $\theta$ range from 0 to $\pi$, and $\phi$ ranges from 0 to $2\pi$. Intuitively this set of coordinates can be deduced by noticing that as one goes up a dimension from the circle, $\mathbb{S}^1$, to the sphere, $\mathbb{S}^2$, that an extra parameter, $\theta$, is introduced which ranges from 0 to $\pi$ in the following manner 
\begin{equation}\label{3rrr}
\begin{aligned}
&x_{\mathbb{S}^2} = \sin \theta x_{\mathbb{S}^1}, \\& y_{\mathbb{S}^2} = \sin \theta y_{\mathbb{S}^1}, \\& z_{\mathbb{S}^2} = \cos \theta.
\end{aligned}
\end{equation}
For the coordinates originally associated with the lower dimensional sphere, a factor of $\sin \theta_{i}$, where $\theta_{i}$ ranges from 0 to $\pi$ is included and represents a new degree of freedom present on the higher dimensional sphere. The new independent Cartesian coordinate which differentiates the higher dimensional embedding space from the one dimension lower space is parameterized by $\cos \theta_{i}$. We see that carrying out this construction from the $\mathbb{S}^2$, parameterized using standard spherical coordinates, to the $\mathbb{S}^3$ results in (\ref{2rrr}). Extending this to arbitrary dimensions results in the following parameterization for a n-sphere with radius $r$,

\begin{equation}\label{4rrr}
\begin{aligned}
x_{1} &=r \cos \left(\theta_{1}\right) \\
x_{2} &=r \sin \left(\theta_{1}\right) \cos \left(\theta_{2}\right) \\
x_{3} &=r \sin \left(\theta_{1}\right) \sin \left(\theta_{2}\right) \cos \left(\theta_{3}\right) \\
& \vdots \\
x_{n-1} &=r \sin \left(\theta_{1}\right) \cdots \sin \left(\theta_{n-2}\right) \cos \left(\phi\right) \\
x_{n} &=r \sin \left(\theta_{1}\right) \cdots \sin \left(\theta_{n-2}\right) \sin \left(\phi\right) .
\end{aligned}
\end{equation}
Because the n-sphere and in turn the n-spheroid can be parameterized in a convenient coordinate system the results in this paper can be generalized to n dimensions. 

Using (\ref{2rrr}) we obtain the following metric representation for a 3D spheroid

\begin{equation}\label{23}
\begin{aligned}
&ds_{sph}^2= \\& \left(b^{2}\sin^{2} \psi +a^{2}\cos^{2} \psi \right) d\psi^2+a^{2}\sin^{2}{\psi} (d\theta^2+\sin^{2}{\theta} d\phi^2).
\end{aligned}
\end{equation}
Using the following change in coordinates inspired by \cite{deng2003conformal}, $w=\int \sqrt{b^{2}\sin^{2} \psi +a^{2}\cos^{2} \psi}d\psi$, we can rewrite (\ref{23}) as

\begin{equation}\label{23rrr}
ds_{sph}^2=dw^{2}+f(w)^{2} (d\theta^2+\sin^{2}{\theta} d\phi^2),
\end{equation}
where $f(w)=a \sin \psi$. This metric can be related conformally to the following metric on $\mathbb{R}^3$

\begin{equation}\label{25rrr}
\begin{aligned}
ds_{flat}^2= (dr^{2}+ r^{2}d\theta^2+r^{2}\sin^{2}{\theta} d\phi^2)
\end{aligned},
\end{equation}
by introducing this functional dependency, $r= e^{g(w)}$, which results in 

\begin{equation}\label{26rrr}
\begin{aligned}
&ds_{flat}^2=\\&e^{2g(w)}\left(\frac{dg}{dw}\right)^{2}\left(dw^{2}+ \left(\frac{1}{\frac{dg}{dw}}\right)^{2}\left(d\theta^2+\sin^{2}{\theta} d\phi^2\right)\right).
\end{aligned}
\end{equation}
We can now set 
\begin{equation}\label{27rrr}
\left(\frac{1}{\frac{dg}{dw}}\right)=a \sin \psi(w)
\end{equation}
and obtain $g(w)=\int\frac{1}{a}\csc \psi dw$. Going back to the original coordinate transformation we applied to (\ref{23}), we can rewrite $g(w)$ as 

\begin{equation}\label{28rrr}
\begin{aligned}
g(w)&=\int \frac{1}{a}\csc \psi \sqrt{b^{2}\sin^{2} \psi  +a^{2}\cos^{2} \psi}d\psi \\& =\int\sqrt{\frac{b^2}{a^2}+\cot^{2} \psi}d\psi. 
\end{aligned}
\end{equation}
By doing so we recover the following metric which is conformal to (\ref{23rrr})

\begin{equation}\label{29rrr}
ds_{flat}^2=\frac{e^{2g(w)}}{a^{2}\sin^{2} \psi}\left(dw^{2}+f(w)^{2} (d\theta^2+\sin^{2}{\theta} d\phi^2)\right).
\end{equation}
By comparing (\ref{29rrr}) to (\ref{23rrr}) we can deduce that the Weyl factor of our 3D spheroid is 
\begin{equation}\label{30rrr}
\Omega_{spheroid}\left(x_{i}\right)=a\sin \psi{e^{-\int\sqrt{\frac{b^2}{a^2}+\cot^{2} \psi}d\psi}}.
\end{equation}
For the case of $\mathbb{S}^{3}$ when $b=a=1$ this reduces to 
\begin{equation}\label{31rrr}
\Omega_{\mathbb{S}^{3}}\left(x_{i}\right)=2\cos^{2} \frac{\psi}{2}.
\end{equation}
Going back to how we parameterized $r$ in (\ref{25rrr}) and setting $b=a=1$ we obtain the following mapping between a point on $\mathbb{S}^{3}$ and a point in $\mathbb{R}^{3}$
\begin{equation}\label{32rrr}
\begin{aligned}
& z=\tan \left(\frac{\psi }{2}\right)\cos{\theta}, \\& y=\tan \left(\frac{\psi }{2}\right)\sin{\theta} \sin{\phi}, \\& x= \tan \left(\frac{\psi }{2}\right)\sin{\theta} \cos{\phi} .
 \end{aligned}
\end{equation}

\begin{figure}[ht!]
\centering
\begin{subfigure}{.4\textwidth}
 \centering
 \includegraphics[scale=.6]{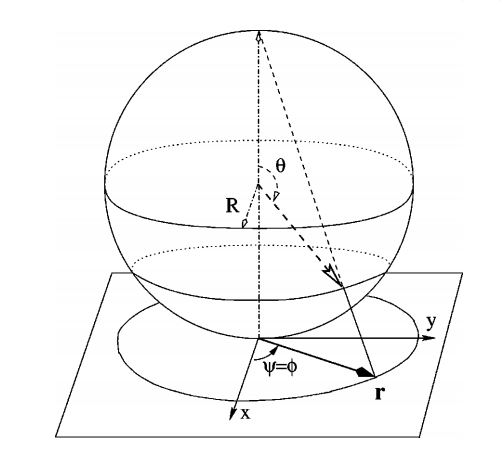}
 \caption{}
 \label{fig 1a}
\end{subfigure}
\caption{Illustration of the conformal mapping between an infinite plane and $\mathbb{S}^{2}$ with the polar angle shifted in the following manner, $\theta \rightarrow \pi-\theta$, compared to what we have in (\ref{31rrr}). This imagine originally appeared in \cite{deng2003conformal}}
\label{fig 1}
\end{figure}

From a geometric perspective our conformal mapping of $\mathbb{S}^{3}$ to $\mathbb{R}^{3}$ is the exact higher dimensional analogue of the standard stereographic projection commonly performed from $\mathbb{S}^{2}$ to $\mathbb{R}^{2}$. For the general case this mapping can be accomplished by placing a $\mathbb{S}^{n}$ on $\mathbb{R}^{n}$ with its south pole centered on the origin of $\mathbb{R}^{n}$ and drawing lines from the north pole which intersect both $\mathbb{S}^{n}$ and $\mathbb{R}^{n}$. Each one of those lines are oriented by a set of n angles and their intersection with $\mathbb{S}^{n}$ and $\mathbb{R}^{n}$ provides a one to one mapping between $\mathbb{S}^{n}$ and $\mathbb{R}^{n}$. The north pole itself cannot be mapped using only a single cover because in the limiting case the line that would intersect the north pole becomes parallel to $\mathbb{R}^{n}$ and hence never intersects it. That is why $\mathbb{S}^{n}$ can be thought of as a one point compactification of $\mathbb{R}^{n}$. A picture of this stereographic projection is provided in Fig.~\ref{fig 1}.

\subsection{Conformal Invariance of $R \cross \mathbb{S}^2$}

As we previously mentioned, in the limit, $b \rightarrow \infty$, (\ref{1rrr}) gives the equation for a 3D cylinder which is topologically equivalent to $R \cross \mathbb{S}^2$. One way of seeing this is to examine what happens to (\ref{1rrr}) as $ b \rightarrow \infty$ and  $-\infty < w < \infty$. In the limit of, $ b \rightarrow \infty$, and $w$ being finite, (\ref{1rrr}) reduces to 

\begin{equation}\label{1zzz}
\frac{x^{2}}{a^{2}}+\frac{y^{2}}{a^{2}}+\frac{z^{2}}{a^{2}}=1.
\end{equation}
Despite the disappearance of $w$ the manifolds which (\ref{1zzz}) admits are still embedded in $\mathbb{R}^{4}$. Thus $w$ can be parameterized independently of x, y, and z in an arbitrary fashion. Because we wish to recover $R \cross \mathbb{S}^2$ we will set $w=t$, where $t$ ranges from $\left(-\infty,\infty\right)$ and parameterize x, y, and z using standard spherical coordinates. Doing so results in the following metric 

\begin{equation}\label{2zzz}
\begin{aligned}
& ds_{3-cyl}^2=dx^{2}+dy^{2}+dz^{2}+dw^{2}, \\& ds_{3-cyl}^2= dt^2+a^{2} (d\theta^2+\sin^{2}{\theta} d\phi^2),
\end{aligned}
\end{equation}
where $a$ is the radius of $\mathbb{S}^2$.
Using (\ref{4rrr}), this metric can be generalized to $R \cross \mathbb{S}^n$
\begin{equation}\label{3zzz}
\begin{aligned}
ds_{n-cyl}^2= dt^2+d\Omega^{2}_{n-1},
\end{aligned}
\end{equation}
where $d\Omega^{2}_{n-1}$ is the metric of $\mathbb{S}^{n-1}$ which can be obtained for any n using (\ref{4rrr}). As the reader can verify $dt^2+a^{2} (d\theta^2+\sin^{2}{\theta} d\phi^2)$ is conformally related to $(dr^{2}+ r^{2}d\theta^2+r^{2}\sin^{2}{\theta} d\phi^2)$ through the following Weyl factor 

\begin{equation}\label{4zzz}
\begin{aligned}
\Omega_{3-cyl}^{2}=e^{-2t/a},
\end{aligned}
\end{equation}
generated by defining $r$ as 
\begin{equation}\label{5zzz}
\begin{aligned}
r=ae^{\frac{t}{a}}.
\end{aligned}
\end{equation}
It should be mentioned that setting $b$ equal to a positive, real, finite number does not result in the geometry of a finite length $ \mathbb{R} \cross \mathbb{S}^2$. Rather the resultant geometry is a 3D ellipsoid. Only in the limit as $b\rightarrow \infty$ is $ \mathbb{R} \cross \mathbb{S}^2$ realized.

Because $ \mathbb{R} \cross \mathbb{S}^2$ isn't a compact manifold using integration to find the Binder cumulant (\ref{34}) isn't trivial. To obtain an estimate of the Binder cumulant for this non-compact geometry one can perform a single point compactification of $ \mathbb{R} \cross \mathbb{S}^2$ which results in $ \mathbb{S}^1 \cross \mathbb{S}^2$ and define a lattice field theory on that manifold. Once a lattice field theory is defined on  $ \mathbb{S}^1 \cross \mathbb{S}^2$ one can perform a Monte Carlo simulation to compute an estimate of the Binder cumulant on that compactified geometry as was done in \cite{Brower:2020jqj}. When this compactification is done the manifold is defined by two radii $r1$ and $r2$, where $r1$ denotes the radius of $\mathbb{S}^1$ and, $r2$ is the radius of $\mathbb{S}^2$. In the limit as $r1 \rightarrow \infty$, $ \mathbb{S}^1 \cross \mathbb{S}^2$ approaches $ \mathbb{R} \cross \mathbb{S}^2$. Therefore if one has a method, such as the QFE, which allows them to formulate lattice field theories on curved manifolds they can study the critical 3D Ising model on  $ \mathbb{S}^1 \cross \mathbb{S}^2$ as $r1 \rightarrow \infty$ and observe what the Binder cumulant approaches.

\subsection{Conformal Invariance of 3-Ball}

Before we move on to calculating the Binder cumulant on $\mathbb{S}^{3}$ it is prudent to talk about how one would do the corresponding calculation for the 3-ball. The 3-ball is imminently related to $\mathbb{S}^{3}$. One can construct $\mathbb{S}^{3}$ by superimposing two 3-balls and defining an equivalence class which identifies all of the points which make up the shared boundaries of this superimposed ball. In other words, the points which make up the boundary of the two superimposed 3-balls, which are two 2-spheres, are identified as a single point. This identification of the boundary with a single point can be realized by projecting out the boundary of this superimposed 3-ball into a higher dimensional space in which the boundary points of the 2-spheres merge at the north pole; thus resulting in a newly formed $\mathbb{S}^{3}$. 

Mathematically we see a hint of this construction by setting, $b=0$, and obtaining the following r from our earlier coordinate transformation 

\begin{equation}\label{33rrr}
\begin{aligned}
& r=\lim_{b\to 0}e^{\int\sqrt{\frac{b}{a} +\cot^{2} \psi} d\psi}=\sin \psi.
 \end{aligned}
\end{equation}
As it can be seen when $b \to 0$ our $r$ doesn't cover the whole plane $\mathbb{R}^{3}$ because $r$ only ranges from 0 to 1. This is because these coordinates only map points located on the lower hemisphere and excludes points located on our 3-ball's upper hemisphere. This is an artifact of $\mathbb{S}^{3}$ being a construction of two superimposed 3-balls with their shared boundaries being glued together at a single point. The total mapping can be realized by noticing that we could have defined our functional dependency right above (\ref{26rrr}) as $r= e^{-g(w)}$. Such a functional dependency would have allowed us to obtain a different Weyl factor, $\Omega(x_{i})$, which nonetheless yields the exact same Binder cumulant for the case of $\mathbb{S}^{3}$ as the one we previously calculated. However it would also result in the following radius 

\begin{equation}\label{34rrr}
\begin{aligned}
& r=\lim_{b\to 0}e^{-\int\sqrt{\frac{b}{a} +\cot^{2} \psi} d\psi}=\csc \psi.
 \end{aligned}
\end{equation}
which ranges from 1 to $\infty$. Thus in order to compute the two and four-point functions for the superimposed 3-ball one must differentiate between points on the northern hemisphere and points on the southern hemisphere because they both have different radial coordinates, (\ref{33rrr}) and (\ref{34rrr}). The resulting calculation involves group averaging the four-point function over all of the different combination of points that can be on one hemisphere and the opposite hemisphere. More information of this group averaging for the 2-ball which can be extrapolated to the 3-ball can be found in \cite{deng2003conformal}. 

Preliminary results obtained using the five operators reported in table 1 of \cite{Komargodski:2016auf} and setting $\frac{b}{a}=10^{-5}$ yields, $U_{4}=0.38703 \pm .0.00570$, which suggests that the Binder cumulant for these two superimposed 3-balls is similar to the Binder cumulant for $\mathbb{S}^{3}$ as we show in section IV. This similarity between the Binder cumulant of the superimposed 3-ball and $\mathbb{S}^{3}$ are in concord with the similarity found in \cite{deng2003conformal} between the Binder cumulant of the superimposed 2-ball(disc) and $\mathbb{S}^{2}$.

For the related case of the interior of a single, non-superimposed sphere that was considered in \cite{cosme2015conformal} one has to take into account the boundary. This is done by studying the boundary conformal field theory(BCFT)\cite{cardy1984conformal,cardy1996scaling} which has its own scaling dimensions and operators associated with it. For the case with a free boundary on $\mathbb{S}^{2}$ the scaling dimension of the relevant operator was found to be \cite{diehl1998massive,deng2005surface,hasenbusch2011thermodynamic,gliozzi2015boundary} $\Delta_{\tilde{\sigma}}=1.276(2)$. Thus in computing the Binder cumulant of a 3-ball with a boundary one must differentiate between pairs of points which are either both on the boundary, inside the bulk or where one point is inside the bulk and the other is on the boundary. The need to differentiate the location of points on the two aforementioned manifolds is a similarity that the two-superimposed 3-balls and the single 3-ball with boundary share with each other. 

Using the known values of the scaling dimensions on the boundary of a 3-ball and its interior allows one in theory to use the formalism we presented in this paper to compute an estimate of the fourth-order Binder cumulant. An estimate of the fourth-order Binder cumulant for the critical 3D Ising model on a 3-ball with a boundary computed by integrating its two and four-point functions can be compared with the estimate obtained in \cite{cosme2015conformal}. Comparing this hypothetical estimate to what was obtained in \cite{cosme2015conformal} can increase our understanding of how to study BCFTs via numerical simulations. In the future we plan to do the calculation we outlined in this section in the limit when, $\frac{b}{a}=0$, and perform a comparison of the value of the Binder cumulant obtained from direct integration to \cite{cosme2015conformal}. Furthermore in the future the QFE can be extended to apply to quantum field theories formulated on curved manifolds with boundaries. 

\section{The Four-Point Function for the Critical 3d Ising Model}

The four-point function for the critical 3D Ising model whose form is restricted by conformal symmetry is given below 

\begin{equation}\label{18}
\left\langle\phi(x_{1}) \phi(x_{2})  \phi(x_{3})  \phi(x_{4}) \right\rangle_{flat}=\frac{g(u, v)}{\left|x_{1}-x_{2}\right|^{2 \Delta_{\sigma}}\left|x_{3}-x_{4}\right|^{2 \Delta_{\sigma}}},
\end{equation}
where $x_{i}$ is a point in $\mathbb{R}^3$ and $u$, and $v$ are the following conformally invariant cross ratios 

\begin{equation}\label{19}
\begin{aligned}
&u=\frac{\left(x_{12}^{2} x_{34}^{2}\right)} {\left(x_{13}^{2} x_{24}^{2}\right)},\\&v=\frac{\left(x_{32}^{2} x_{14}^{2}\right)} {\left(x_{13}^{2} x_{24}^{2}\right)}.
\end{aligned}
\end{equation}
For our four-point function (\ref{18}), $\Delta_{\sigma}$ is the scaling dimension of $\phi(x_{i})$ and its value \cite{kos2016precision} estimated by the conformal bootstrap is $\Delta_{\sigma}= 0.5181489(10)$. The numerator of (\ref{18}), $g(u, v)$, can be expressed as a OPE in terms of conformal blocks 

\begin{equation}\label{20}
g(r, \eta)=1+\sum_{\mathscr{O} \in \sigma \times \sigma} C_{\sigma \sigma \mathscr{O}}^{2} g_{\Delta_{\mathscr{O}}, \ell_{\mathscr{O}}}(r, \eta),
\end{equation}
where 
\begin{equation}\label{21}
\begin{aligned}
&r=\sqrt{\frac{z \bar{z}}{\left(\sqrt{1-z}+1\right)^2 \left(\sqrt{1-\bar{z}}+1\right)^2}} \\&\eta=\frac{\frac{z}{\left(\sqrt{1-z}+1\right)^2}+\frac{\bar{z}}{\left(\sqrt{1-\bar{z}}+1\right)^2}}{2 \left(\frac{z \bar{z}}{\left(\sqrt{1-z}+1\right)^2 \left(\sqrt{1-\bar{z}}+1\right)^2}\right)^{\frac{1}{2}}}\\& u=z\bar{z} \\& v= (1-z)(1-\bar{z}).
\end{aligned}
\end{equation}
More information on the physical meaning behind these coordinates (r, $\eta$, $z$, $\bar{z}$) can be found in \cite{rychkov2017non,kos2014bootstrapping}. 

The sum in (\ref{20}) is over the operators (excluding the unit operator) that are present in the $\phi(x_{i})\times \phi(x_{j})$ OPE of dimension $\Delta_{\mathscr{O}}$ and spin $\ell_{\mathscr{O}}$. The scale dimensions and spins for these operators are provided in table 2 of \cite{Simmons-Duffin:2016wlq}. To evaluate (\ref{20}) we use following recursion relation which was first reported in \cite{zkos2014bootstrapping}

\begin{equation}\label{22}
\begin{aligned}
&(4r)^{\Delta}h_{\Delta, \ell}(r, \eta) \equiv  g_{\Delta, \ell}(r, \eta) \\&
h_{\Delta, \ell}(r, \eta)=h_{\ell}^{(\infty)}(r, \eta)+\sum_{k} \frac{c_{1} r^{n_{1}}}{\Delta-\Delta_{1}} h_{\Delta_{1}+n_{1}, \ell_{1}}(r, \eta) \\& +\sum_{k} \frac{c_{2} r^{n_{2}}}{\Delta-\Delta_{2}} h_{\Delta_{2}+n_{2}, \ell_{2}}(r, \eta) +\sum_{k} \frac{c_{3} r^{n_{3}}}{\Delta-\Delta_{3}} h_{\Delta_{3}+n_{3}, \ell_{3}}(r, \eta)
\end{aligned}
\end{equation}
where information on what $h_{\ell}^{(\infty)}(r, \eta)$, $c_{i}$, $n_{i}$, $\ell_{i}$, and $\Delta_{i}$ are is described in \cite{kos2014bootstrapping}.

This recursion relation converges quickly and is easy to evaluate using a computer algebra system like Mathematica. For our purposes we evaluate $h_{\Delta, \ell}$ to 12th order in $r$ where $h_{\ell}^{(\infty)}$ is not expanded in terms of r. The result of using this recursion relation (\ref{22}) to compute $h(r, \eta)$ up to nth order should be a nth order polynomial in r whose coefficients include $h_{\ell}^{(\infty)}(r, \eta)$. In the appendix, there will be a Mathematica code that evaluates (\ref{20}) as a sum over the operators present in table 1 of    \cite{komargodski2017random} up to any order in $r$. The code can be easily modified to include the eleven terms which appear in table 2 of \cite{Simmons-Duffin:2016wlq}.

\section{Binder Cumulant Estimate}

Using the Weyl factor, $\Omega\left(x_{i}\right)= 2\cos ^2\left(\frac{\psi_{i} }{2}\right)$, we obtained earlier we can map the two and four-point functions of the critical 3D Ising model from $\mathbb{R}^3$ to $\mathbb{S}^3$ as is shown below in 
\begin{equation}\label{28}
\begin{aligned}
&\left\langle\phi\left(x_{1}\right) \phi\left(x_{2}\right)\right\rangle_{g_{u v}}=\frac{1}{\Omega\left(x_{1}\right)^{\Delta_{\sigma}}} \frac{1}{\Omega\left(x_{2}\right)^{\Delta_{\sigma}}}\left\langle\phi\left(x_{1}\right) \phi\left(x_{2}\right)\right\rangle_{\text {flat }} \\&\left\langle\phi\left(x_{1}\right) \phi\left(x_{2}\right) \phi\left(x_{3}\right) \phi\left(x_{4}\right)\right\rangle_{g_{u v}}\\&=\frac{1}{\Omega\left(x_{1}\right)^{\Delta_{\sigma}}} \ldots\frac{1}{\Omega\left(x_{4}\right)^{\Delta_{\sigma}}}\left\langle\phi\left(x_{1}\right) \ldots \phi\left(x_{4}\right)\right\rangle_{\text {flat }}.
\end{aligned}
\end{equation}
We will use these variables in (\ref{32rrr}) to construct our two and four-point functions

As it can be seen below conformal symmetry greatly restricts the form of the following two-point function 
\begin{equation}\label{31}
\left\langle\phi\left(x_{1}\right) \phi\left(x_{2}\right)\right\rangle_{flat}=\frac{1}{x_{12}^{2\Delta}},
\end{equation}
where $x_{ij}$ =$|x_i-x_j|$. 
The two quantities we need to find by integrating our two and four-point functions over $\mathbb{S}^3$ in order to obtain our fourth-order Binder cumulant are the following magnetization densities 
\begin{equation}\label{32}
\begin{array}{l}
\left\langle\sigma^{2}\right\rangle=\rho^{2} \int \mathrm{d} S_{1} \mathrm{~d} S_{2} \left\langle\phi\left(x_{1}\right) \phi\left(x_{2}\right)\right\rangle_{g_{u v}}, \quad  \\
\left\langle\sigma^{4}\right\rangle=\rho^{4} \int \mathrm{d} S_{1} \cdots \mathrm{d} S_{4}\left\langle\phi\left(x_{1}\right) \phi\left(x_{2}\right) \phi\left(x_{3}\right) \phi\left(x_{4}\right)\right\rangle_{g_{u v}},
\end{array}
\end{equation}
where $\rho$ is the areal density of the spins, and $\mathrm{d} S_{i}$ represents the number of spins in an infinitesimal area. For the $\mathbb{S}^3$, $\rho$, and $\mathrm{d} S_{i}$ can respectively be expressed as $\frac{1}{2\pi^{2}}$ and $\sin^{2}{\psi_{i}}\sin{\theta_{i}}d\psi_{i} d\theta_{i} d\phi_{i}$. We can reduce the computational cost of integrating our two and four-point functions by taking advantage of the SO(4) symmetry of $\mathbb{S}^3$. Because SO(4) has six generators, which corresponds to six independent rotations, we can rotate our $\mathbb{S}^3$ in such a way that some of the angles that we would normally need to integrate over are fixed; thus reducing the computational cost of our multidimensional Monte Carlo integration. This can be seen because SO(4) is locally isomorphic to $SO(3) \otimes SO(3)$. Thus there are six independent rotations for $\mathbb{S}^3$ three for each SO(3). This can be seen by noticing that the Lie Algebra of SO(4) can be represented as two copies of the Lie Algebra of SO(3). 

The key for efficiently evaluating these integrals (\ref{32}) is to use the SO(4) symmetry of $\mathbb{S}^3$. For the two-point function we can naively evaluate a 6th dimensional integral over these points on its surface, $ (\psi_1, \theta_1, \phi_1)$ and $ (\psi_2, \theta_2, \phi_2)$. By rotating $\mathbb{S}^3$ we can set the first point to be $(0,0,0)$ and the second point to be $(\psi_2,0,0)$. This results in the following integral for the 2nd order magnetization density. 

\begin{equation}\label{33}
\left\langle\sigma^{2}\right\rangle=\int_0^{\pi } \frac{\left(\left(2 \pi ^2\right) (4 \pi ) \sin ^2\left(\psi
   _2\right)\right) \left(\frac{1}{2 \pi ^2}\right)^2}{\left( 2 \cos ^2\left(\frac{\psi _2}{2}\right)\right){}
   \left(\frac{\sin ^2\left(\psi _2\right)}{\left(1+\cos \left(\psi
   _2\right)\right){}^2}\right)^{0.518149}} \, d\psi _2
\end{equation}
which yields from Mathematica's INTEGRATE function
\begin{equation}\label{1ext}
\left\langle\sigma^{2}\right\rangle=0.847359
\end{equation}

The four-point function can be ultimately expressed using the following coordinates $(\psi_1, \theta_1, \phi_1)$, $(\psi_2, \theta_2, \phi_2)$, $(\psi_3, \theta_3, \phi_3)$, $(\psi_4, \theta_4, \phi_4)$. Using the SO(4) group we can reduce the dimensionality of our integral for $\left\langle\sigma^{4}\right\rangle$ from twelve to six by fixing the following coordinates  $ (0,0,0),(\psi_2,0,0),(\psi_3, \theta_3 ,0), (\psi_4, \theta_4, \phi_4)$. Using Mathematica NINTEGRATE, we preformed 10,000 Monte Carlo evaluations and obtained the following estimate of the fourth-order magnetization and its associated statistical error
\begin{equation}\label{2ext}
\left\langle\sigma^{4}\right\rangle=1.59083 \pm 0.00016.
\end{equation}

We now have all that we need to compute an estimate of the fourth-order Binder cumulant. 
\begin{equation}\label{34}
U_4= \frac{3}{2} \left(1-\frac{1}{3} \frac{\left\langle\sigma^{4}\right\rangle}{\left\langle\sigma^{2}\right\rangle^{2}}\right)
\end{equation}
\begin{equation}\label{3ext}
U_4=0.39220 \pm 0.00011.
\end{equation}
For now we exclude sources of error orientating from uncertainty inherent to the CFT data obtained through the bootstrap.

This result must be understood within the context of the OPE representation of the four-point function. Infinitely many operators of varying scaling dimension and spin exist which must be summed in order to obtain an exact expression for the four-point function of the critical 3D Ising model. The finite number of operators whose CFT data has been obtained from the bootstrap are the leading order operators which contribute the most to the four-point function.  However, because we only took into account data pertaining to eleven of those operators a systematic error will be present in our calculation as a result of us not being able to integrate the exact four-point function. The remainder of the operators posses a higher spin and/or scaling dimension. Thus their inclusion would allow us to more accurately compute the four point function when the points are very close to each other. 

The range of this systematic error can be estimated by taking the difference between the Binder cumulant computed using the eleven operators listed in table 2 of \cite{Simmons-Duffin:2016wlq} and the resultant cumultant one obtains if they use only ten of the operators. There is no definitive answer for which ten operators one should include. Thus we performed this estimate using two similar, but different methodologies. The first methodology involves computing a sequence of Binder cumulants as operators are added to the OPE in the order of their scaling dimension. This means first computing the Binder cumulant while only including the operator with the lowest scaling dimension, $\epsilon$, in the OPE representation of the four-point function and than including the operator with the next lowest scaling dimension, $T_{\mu, \nu}$, until ten of the eleven operators are included in the four-point function. The next methodology is similar except operators are included sequentially into the OPE in order of increasing spin. 

In terms of increasing scaling dimension and spin, one respectively obtains the following Binder cumulants and potential estimates for the systemic error 
\begin{equation}\label{100ext}
\begin{aligned}
& U_{Scaling}=0.39216 \pm 0.00011 \\&
\Delta U_{4-Scaling}=U_{4}-U_{Scaling}=0.00004,
\end{aligned}
\end{equation}

\begin{equation}\label{99ext}
\begin{aligned}
& U_{Spin}=0.39165\pm 0.00011 \\&
\Delta U_{4-Spin}=U_{4}-U_{Spin}=0.00141.
\end{aligned}
\end{equation}

One way to interpret the magnitude of the systematic errors that we obtained using bootstrap data is to compare it to the magnitude of the systematic error generated by doing the analogous calculation using the QFE. A direct comparison of such nature cannot be done at the moment because we haven't had a chance to apply the QFE to $\phi^{4}$ theory on $\mathbb{S}^3$. However it is reasonable to expect that the relative error that we will obtain when we do the aforementioned calculation will be similar to the relative error obtained for the analogous calculation \cite{mohamed2018numerical,gasbarro2018studies} on $\mathbb{S}^2$ which has already been done. 

Below are the statistical(58) and systematic(90) errors for an estimate of the fourth-order Binder cumulant of $\phi^{4}$ theory at its Wilson-Fisher conformal fixed point on $\mathbb{S}^2$ computed using the QFE (Monte Carlo Values) \cite{mohamed2018numerical,gasbarro2018studies} and direct integration (Analytic CFT Values).

\begin{equation}\label{1est}
\begin{array}{ll}
\text { Monte Carlo Values: } & U_{4, c r}=0.85020(58)(90) \\
\text { Analytic CFT Values: } & U_{4}^{*}=0.8510207(63).
\end{array}
\end{equation}
If we compute the relative error of the above QFE result we obtain
\begin{equation}\label{2est}
\delta U_{4, c r}= \frac{9 \cross 10^{-4}}{0.85020} \approx 10^{-3}.
\end{equation}
This is a reasonable estimate of the systematic error that we expect our QFE calculation on $\mathbb{S}^3$ to yield. Therefore in order to interpret the magnitude of the relative error that the bootstrap presently yields for the fourth-Binder cumulant of the critical 3D Ising model we should compare it to (\ref{2est}). 

\begin{equation}\label{3est}
\delta U_{4, scaling-dim}  \approx 9 \cross 10^{-5},
\end{equation}
\begin{equation}\label{4est}
\delta U_{4, spin} \approx 3.6 \cross 10^{-3}.
\end{equation}
If the systematic error in our calculation is closer to (\ref{3est}) that would suggest that the current CFT data that we have from the bootstrap is enough to compute an accurate estimate of the four-point function relative to the implementation of the QFE that was used in \cite{Brower:2018szu}. However if the systematic error is much closer to (\ref{4est}), that would indicate that the current bootstrap results {aren't} enough to match the accuracy of the QFE and that additional data on higher order operators is needed so that the accuracy of the two calculations can be in agreement with each other.

To demonstrate the convergence of the Binder cumulant as we add terms to the four-point function we show in figure 2 the Binder cumulant as a function of these operators for both methodologies. We used Monte-Carlo integration and 1,000 iterations to compute each Binder cumulant. We are confident that our range is representative of the systematic error because it is evident that as we include operators in the OPE that our results are converging to a definitive value. The difference between preceding cumulants have a tendency to shrink as we include more operators, thus the inclusion of more operators will allow us to further minimize the systematic error. Thus in order to compute a more accurate estimate of the fourth-Binder cumulant using integration we need additional bootstrap results for higher order operators and to include more Monte Carlo iterations.  

Figure 2b shows an interesting phenomenon. Excluding when only the operator with both the lowest scaling dimension and spin is included, the inclusion of an operator with higher spin results in the value of $U_{4}$ jumping. When the highest operator included in the four-point function has spin 2 the values of $U_{4}$ varies little as additional spin 2 operators are included. It is only when spin 4 operators are included that we see such a jump and again see that the value of $U_{4}$ varies very little when additional spin 4 operators are included. We see a similar jump when we include a spin 6 operator. The jump though decreases in magnitude as higher and higher spin operators are included. This suggests that the value of the Binder cumulant approaches some definitive number as we increase the operators in the OPE representation of the four point function. The origin of this phenomenon deserves to be investigated.

As a check that our procedure for evaluating the four-point function on $\mathbb{S}^3$  is correct we calculated the Binder cumulant for the free theory. The correlation functions for a free CFT are given in \cite{guerrieri2016free} and its Binder cumulant should be zero. Using Monte Carlo integration, 10,000 iterations, and accuracy goal 15, we calculated 
\begin{equation}\label{4ext}
U_4= -1.9176 \times  10^{-6} \pm 4.7357 \times  10^{-5}.
\end{equation}
Our result is very comfortably within the range of the expected result of 0 for the Binder cumulant. We hope to check our results for the Binder
cumulant of the critical 3D Ising model on $\mathbb{S}^3$ using the QFE in the near future.

\section{Concluding Remarks}
Using the data of the critical 3D Ising model computed using the conformal bootstrap method, we integrated the approximate two and four-point functions to obtain an estimate of the fourth-order Binder cumulant. We also showed how this approach could be used to estimate the Binder cumulants for the 3-ball and other 3D spheroids.  Our approach is an extension of the work by Deng and Blote \cite{deng2003conformal} to three dimensions and we showed how it could be extended further to higher dimensional spheroids.  The immediate application of our result is to compare this estimate of the Binder cumulant with one computed in an upcoming calculation of $\phi^{4}$ theory on $\mathbb{S}^3$ using quantum finite elements (QFE). A favorable comparison of the two methods would give us further confidence
that QFE is a correct framework for computing non-perturbative quantum field
theories on curved manifolds.

\onecolumngrid\
\begin{figure}
\centering
\begin{subfigure}{.4\textwidth}
 \centering
 \includegraphics[scale=.24]{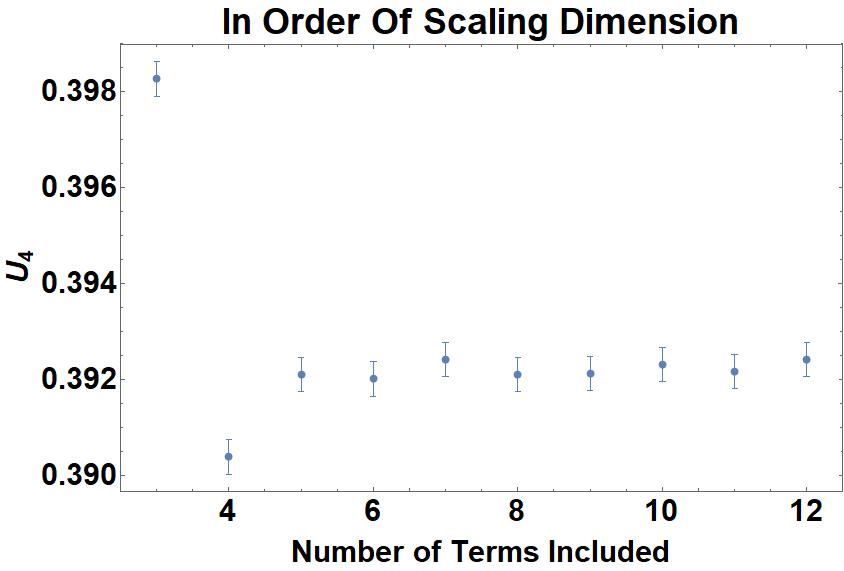}
 \caption{Binder cumulant plot in order of scaling dimension }
 \label{fig 8a}
\end{subfigure}%
\begin{subfigure}{.4\textwidth}
 \centering
 \includegraphics[scale=.23]{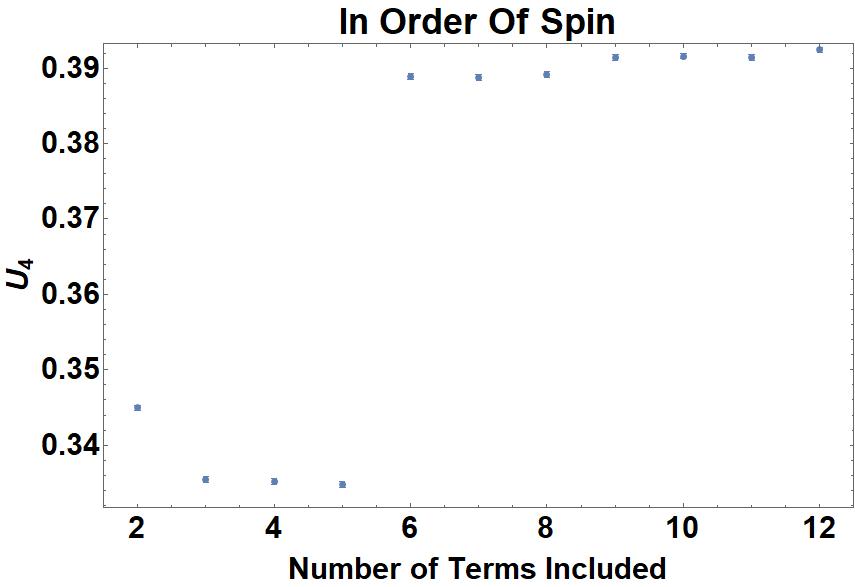}
 \caption{Binder cumulant plot in order of increasing spin }
 \label{fig 8b}
\end{subfigure}
\caption{Plots of the Binder cumultant as we include operators into the OPE of the four-point function in order of their scaling dimension and spin. We count the unit operator as the first term included. We started with the 3rd operator plot in order of scaling dimension. The statistical error bars don't show in the spin graph because of the large difference in values of the Binder cumulant when only spin zero operator are included versus when spin $2>$ operators are included. The statistical errors for all of the points in our spin graph are around $\approx .0004$.  }
\label{fig 8}
\end{figure}
\twocolumngrid\

\begin{acknowledgments}
Both authors thank Richard Brower for his immense contributions towards the development of the QFE. In addition we thank him for the intellectually stimulating conversations that led us down the road to doing the calculation we presented in this manuscript.  We acknowledge support from the United States Department of Energy through grant number DE-SC0019061. 
\end{acknowledgments}
\onecolumngrid\
\appendix

\section{Appendixes}

Using the following code and the CFT data reported for the five operators included in table 1 of \cite{komargodski2017random} we performed 100,000 Monte Carlo evaluations with our accuracy goal set to 5 and calculated  $\left\langle\sigma^{4}\right\rangle$ to be $1.591463$ with an error of $0.000050$. This allows us to obtain the following estimate of the fourth-order Binder cumulant 
\begin{equation}
U_4=0.391765 \pm 0.000035.
\end{equation}
The code below can be easily modified to include the additional CFT data \cite{Simmons-Duffin:2016wlq} that we used to compute (\ref{3ext}).

This code defines the $h_{\Delta, \ell}(r, \eta)$ which we wish to calculate through the recursion relation given in (\ref{22}). The only input for the user is "n", which is the order one wishes to compute the recursion relation up to.   

\begin{verbatim}
H[\[CapitalDelta]\[CapitalDelta]_, 
  LL_] := {n = 12;, 
   recursion[a_] := Normal[Series[a /. hh -> h, {r, 0, n}]], 
   Nest[recursion, {h[\[CapitalDelta]_, 
        L_] := ((LegendreP[L, \[Eta]]/(
          Sqrt[1 - rr^2] Sqrt[-4 \[Eta]^2 rr^2 + (1 + rr^2)^2]) + 
          Sum[((-((
                2^(1 - 4 k)
                  k ((2 k)!)^2 Pochhammer[1 + L, 
                  2 k])/((k!)^4 Pochhammer[1/2 + L, 2 k])))*(r)^(2*
                  k)/(\[CapitalDelta] - (1 - L - 2*k)))*
            hh[1 - L, L + 2*k], {k, 1, n/2}]
          + 
          Sum[((-((
                k (1/2 - k + L) Pochhammer[1/2, k]^2 Pochhammer[
                  1/4 (3 - 2 k + 2 L), 
                  k]^2)/((1/2 + k + L) (k!)^2 Pochhammer[
                  1/4 (1 - 2 k + 2 L), k]^2)))*(r)^(2*
                  k)/(\[CapitalDelta] - ((3/2) - k)))*
            hh[(3/2) + k, L], {k, 1, n/2}] + 
          Sum[((-((
                2^(1 - 4 k)
                  k ((2 k)!)^2 Pochhammer[1 - 2 k + L, 
                  2 k])/((k!)^4 Pochhammer[3/2 - 2 k + L, 
                  2 k])))*(r)^(2*
                  k)/(\[CapitalDelta] - (2 + L - 2*k)))*
            hh[2 + L, L - 2*k], {k, 1, 
            L/2}])), \[CapitalDelta] = \
\[CapitalDelta]\[CapitalDelta];, 
      L = LL;, (((LegendreP[L, \[Eta]]/(
          Sqrt[1 - rr^2] Sqrt[-4 \[Eta]^2 rr^2 + (1 + rr^2)^2]) + 
          Sum[((-((
                2^(1 - 4 k)
                  k ((2 k)!)^2 Pochhammer[1 + L, 
                  2*k])/((k!)^4 Pochhammer[1/2 + L, 2*k])))*(r)^(2*
                  k)/(\[CapitalDelta] - (1 - L - 2*k)))*
            hh[1 - L, L + 2*k], {k, 1, n/2}]
          + 
          Sum[((-((
                k (1/2 - k + L) Pochhammer[1/2, k]^2 Pochhammer[
                  1/4 (3 - 2 k + 2 L), 
                  k]^2)/((1/2 + k + L) (k!)^2 Pochhammer[
                  1/4 (1 - 2 k + 2 L), k]^2)))*(r)^(2*
                  k)/(\[CapitalDelta] - ((3/2) - k)))*
            hh[(3/2) + k, L], {k, 1, n/2}] + 
          Sum[((-((
                2^(1 - 4 k)
                  k ((2 k)!)^2 Pochhammer[1 - 2 k + L, 
                  2 k])/((k!)^4 Pochhammer[3/2 - 2 k + L, 
                  2 k])))*(r)^(2*
                  k)/(\[CapitalDelta] - (2 + L - 2*k)))*
            hh[2 + L, L - 2*k], {k, 1, L/2}])))}[[4]], n/2]}[[3]]
\end{verbatim}

This code is the numerator that appears in the four-point function for the critical 3D Ising  model and is computed in accordance with (\ref{20})

\begin{verbatim}

G = (1 + (H[1.412625, 0]*(((4*r)^(1.412625))*(1.0518537)^2)) + (H[
       3.82966, 0] (((4*r)^(3.82966))*(0.053029)^2)) + (H[3, 
       2]*(((4*r)^(3)) ((0.5181489/Sqrt[0.946539])^2))) + (H[5.509, 
       2] (((4*r)^(5.509))*(0.0172)^2)) + (H[5.02274, 
       4] (((4*r)^(5.02274))*(0.1319)^2)));
\end{verbatim}

This last piece of code plots the four-point function, showing that our computation of the four-point function agrees with \cite{rychkov2017non}.

\begin{verbatim}
{FourPointFunction = (G/(1 + 
            Abs[z]^1.0362978 + (Abs[z]^1.0362978/
               Abs[1 - z])^1.0362978) /. rr -> r /. 
        r -> Abs[
          z/(1 + Sqrt[1 - z])^2] /. \[Eta] -> (z/(1 + Sqrt[1 - z])^2 +
            Conjugate[z]/(1 + Sqrt[1 - Conjugate[z]])^2)/(2*
           Abs[z/(1 + Sqrt[1 - z])^2]) /. z -> x + I*y);, 
  Plot3D[FourPointFunction, {x, -1, .5}, {y, -1, 1}, 
   RegionFunction -> Function[{x, y}, Sqrt[x^2 + y^2] < 1 && x < 1/2],
    PlotRange -> All]}[[2]]
\end{verbatim}
\twocolumngrid\

\bibliography{PRD_Binder}

\end{document}